\documentclass[aps, a4paper,superscriptaddress,12pt,preprintnumbers,floatfix,nofootinbib,amsmath,amssymb]{revtex4}
\usepackage{amstext,amssymb}
\usepackage{amsmath}
\usepackage{graphicx}
\usepackage{dcolumn}
\usepackage[hyperfootnotes=false]{hyperref}
\usepackage{xspace}
\usepackage{braket}
\usepackage{color}
\usepackage{units}
\usepackage{slashed}
\usepackage{graphicx}
\usepackage{bm}
\newcommand {\ignore}[1]{}
 \def\one{\ensuremath{\mathbf{1}}}
 
 \def\three{\ensuremath{\mathbf{3}}}
 \def\threeS{\ensuremath{\mathbf{3^*}}}	
\def\six{\ensuremath{\mathbf{6}}}
\def\sixS{\ensuremath{\mathbf{6^*}}}
\def\eight{\ensuremath{\mathbf{8}}}
 
\def\3331{$\mathrm{SU(3)_c \times SU(3)_L \times SU(3)_R \times U(1)_{X}}$}
 \newcommand{\sm}{Standard  Model } 
\def\vev#1{\left\langle #1\right\rangle}

\def\lfv{lepton flavour violation }

\begin{document}
\title{Towards gauge coupling unification in left-right symmetric $\mathrm{SU(3)_c \times SU(3)_L \times SU(3)_R \times U(1)_{X}}$ theories}

 \author{\hspace*{-1.9cm}Chandan Hati}
 \email{chandan@prl.res.in}
 \affiliation{\hspace*{-0.0cm}Theoretical Physics Division, Physical Research Laboratory, Ahmedabad 380009, India}
 \affiliation{\hspace*{-0.0cm}Discipline of Physics, Indian Institute of Technology Gandhinagar, Ahmedabad 382424, India}
 \author{Sudhanwa Patra}
\email{sudha.astro@gmail.com}
\affiliation{Center of Excellence in Theoretical and Mathematical Sciences,
 Siksha 'O' Anusandhan University, Bhubaneswar-751030, India}

\author{Mario Reig}
\email{mareiglo@alumni.uv.es}
\affiliation{AHEP Group, Institut de F\'{i}sica Corpuscular --
  C.S.I.C./Universitat de Val\`{e}ncia, Parc Cient\'ific de Paterna.\\
 C/ Catedr\'atico Jos\'e Beltr\'an, 2 E-46980 Paterna (Valencia) - Spain}
\author{Jos\'e W.F. Valle}
\email{valle@ific.uv.es}
\affiliation{AHEP Group, Institut de F\'{i}sica Corpuscular --
  C.S.I.C./Universitat de Val\`{e}ncia, Parc Cient\'ific de Paterna.\\
 C/ Catedr\'atico Jos\'e Beltr\'an, 2 E-46980 Paterna (Valencia) - Spain}
\author{C.A. Vaquera-Araujo}
 \email{vaquera@ific.uv.es}
\affiliation{AHEP Group, Institut de F\'{i}sica Corpuscular --
  C.S.I.C./Universitat de Val\`{e}ncia, Parc Cient\'ific de Paterna.\\
 C/ Catedr\'atico Jos\'e Beltr\'an, 2 E-46980 Paterna (Valencia) - Spain}
\begin{abstract}
  We consider the possibility of gauge coupling unification within the
  simplest realizations of the
  $\mathrm{SU(3)_c \times SU(3)_L \times SU(3)_R \times U(1)_{X}}$
  gauge theory.  We present a first exploration of the renormalization
  group equations governing the ``bottom-up'' evolution of the gauge
  couplings in a generic model with free normalization for the
  generators. Interestingly, we find that for a \3331 symmetry
  breaking scale $M_X$ as low as a few TeV one can achieve unification
  in the presence of leptonic octets. We briefly comment on possible
  grand unified theory frameworks which can embed the
  $\mathrm{SU(3)_c \times SU(3)_L \times SU(3)_R \times U(1)_{X}}$
  model as well as possible implications, such as lepton flavour
  violating physics at the LHC.
\end{abstract}
\maketitle

\section{Introduction}

There is no doubt, at least among theorists, that there must be new
physics, unaccounted for by the \sm successfully describing most of the
phenomena in particle physics.
From the observational side, the theory lacks neutrino masses, needed
to describe
oscillations~\cite{Kajita:2016cak,McDonald:2016ixn,Forero:2014bxa}.
It also fails in accounting for the current cosmological puzzles, as
well as the need to ultimately include gravity as part of the
fundamental theory.
From a more aesthetical point of view the \sm does not offer a basic
understanding of many theoretical issues, such as the origin of parity
violation. Indeed, the chiral nature of the weak interactions remains
a mystery at a fundamental level.  As a matter of fact in higher
unified theories it constitutes an arbitrary input, by no means
automatic.
Likewise, the \sm gives no input on the number of fermion families nor
any understanding of the flavor problem -- neither an understanding of
fermion masses themselves, nor of the fermion mixing patterns --
both of which remain mysteries.
And the list goes on~\cite{Valle:2015pba}.

Frustratingly enough, so far there is no strong hint of new physics
from the Large Hadron Collider at CERN, despite the floods of data
already obtained.
No sign yet of supersymmetry, the leading candidate theory in terms of
which to understand theoretical issues such as the consistency of the
electroweak breaking mechanism, now vindicated by the celebrated
discovery of the Higgs boson~\cite{Aad:2012tfa,Chatrchyan:2012xdj}.

Indeed, by establishing the existence of scalar particles in nature,
the discovery of the Higgs boson has paved the way for probing
extensions of the Standard Model (SM) at accelerators.
One of the extensions of the \sm consists of left-right symmetric
schemes which can give a dynamical basis for describing parity
violation in weak interactions and shed light on the origin of
neutrino mass~\cite{mohapatra:1980ia}.
An interesting alternative is the 331 model, which has the special
feature that it is not anomaly free in each generation of fermions,
but becomes anomaly free only when all the three generations of
fermions are included in the theory~\cite{Singer:1980sw}, a feature
which may shed light on the flavor puzzle, since one family of quarks
must be distinguished from the others.

Both approaches provide a solution to neutrino mass problem either
through the seesaw
mechanism~\cite{GellMann:1980vs,yanagida:1979as,mohapatra:1980ia,Schechter:1980gr,Lazarides:1980nt}
or through other alternative neutrino mass generation mechanisms
\cite{Boucenna:2014ela,Boucenna:2014dia}.
Both frameworks may yield new physics that can be observed at the LHC
or the next generation accelerators \cite{Das:2012ii,Queiroz:2016gif,Aaboud:2016cth}. Which alternative is right, if any, we currently do not know.

Recently a realistic theory framework based on the extended
$\mathcal{G}_{3_C 3_L 3_R 1_X}=SU(3)_C \times SU(3)_L \times SU(3)_R
\times U(1)_X$
gauge group has been proposed. It requires the number of families to
match the number of colors, while encompassing the idea of left-right
symmetry. The new framework admits both high \cite{Reig:2016tuk} as
well as low scale~\cite{Reig:2016vtf} realizations of the seesaw
mechanism of neutrino mass generation, and brings in a plethora of
possible new physics at accessible accelerator energies. The low energy 
phenomenology like neutrinoless double beta decay has been explored recently in \cite{Borah:2017inr} 
where fermion masses including light neutrinos are governed by universal seesaw 
with extra vector like fermions and without having scalar bitriplet. 
An attractive feature is that, depending on how the symmetry breaks to
the Standard Model, one recovers either a conventional left-right
symmetric theory, or a 331 symmetry as the ``next" step towards new
physics.

However, as noted in \cite{Deppisch:2016jzl}, the fact that different
multiplets of the
$\mathcal{G}_{3_C 3_L 3_R 1_X}=SU(3)_C \times SU(3)_L \times U(1)_X$
group appear with different multiplicities makes it difficult to unify
the model within Grand Unified Theories (GUTSs) using canonical
routes~\footnote{For this reason possible string completions have been
  suggested in~\cite{Addazi:2016xuh}.}.
Here we perform the first step towards a possible unification of the
gauge couplings within this class of left-right symmetric
theories. This consists of a first exploration of the renormalization
group equations governing the ``bottom-up'' evolution of the gauge
couplings in a generic model with free normalization for the
generators.

The article is organized as follows. In
Section~\ref{sec:model-framework} we discuss the basic structure of
the $\mathcal{G}_{3_C 3_L 3_R 1_X}$, also called 3331 model, for
short. We consider models with sextet 
Higgs scalars to account for neutrino masses.
In Section~\ref{sec:renorm-group-equat-1} we analyze the resulting
renormalization group (RG) running of the gauge couplings in the model
with and without additional octet states, discussing in detail the
necessary conditions for gauge coupling unification.
For simplicity we assume direct breaking to the Standard  Model.
We conclude that, in the presence of octets, unification is possible
while leaving a light ``gauge boson portal'' which may open access to
new physics at collider energies~\cite{Deppisch:2013cya}. The latter
may allow one to probe physics beyond the \sm in a novel way. We also
comment briefly on the possible embedding of the model within a higher
unification group.

\section{Theory framework}
\label{sec:model-framework}

The gauge group of the left-right symmetric model considered for the
present work is given by
\begin{equation}
\mathcal{G}_{3_C 3_L 3_R 1_X}=SU(3)_C \times SU(3)_L \times SU(3)_R \times U(1)_X\,,
\end{equation}
where the electric charge relation is given by the following formula
\begin{equation}{\label{cheqn}}
	Q = T_{3L} + T_{3R} + \beta \left( T_{8L} + T_{8R} \right) + X\,.
\end{equation}
and the fermion assignments are given as
\begin{eqnarray}
 \Psi_{aL} &= \begin{pmatrix}\nu_{aL} \\ \ell^-_{aL} \\ \chi^q_{aL} \end{pmatrix} \, \sim
(\one,\three,\one,\frac{q-1}{3})\, , & 
\Psi_{aR} = \begin{pmatrix}\nu_{aR} \\ \ell^-_{aR} \\ \chi^q_{aR} \end{pmatrix} \, \sim
(\one,\one,\three,\frac{q-1}{3})\, ,\nonumber \\[1mm]
 Q_{\alpha L} &= \begin{pmatrix} d_{\alpha L} \\ u_{\alpha L} \\ J^{-q-1/3}_{\alpha L} \end{pmatrix} \, \sim
(\three,\threeS,\one,\frac{-q}{3})\, , & 
Q_{\alpha R} = \begin{pmatrix} d_{\alpha R} \\ u_{\alpha R} \\ J^{-q-1/3}_{\alpha R} \end{pmatrix} \, \sim
(\three,\one,\threeS,\frac{-q}{3})\,, \nonumber \\[1mm]
 Q_{3 L} &= \begin{pmatrix} u_{3 L} \\ d_{3 L} \\ J^{q+2/3}_{3 L} \end{pmatrix} \, \sim
(\three,\three,\mathbf{1},\frac{q+1}{3})\, , & 
Q_{3 R} = \begin{pmatrix} u_{3 R} \\ d_{3 R} \\ J^{q+2/3}_{3 R} \end{pmatrix} \, \sim
(\three,\one,\three,\frac{q+1}{3})\,.
\end{eqnarray}
Notice that the electric charge of third component of lepton triplet is related to
$\beta$ parameter in the following way
\begin{equation}
	\beta = -\frac{2 q+1}{\sqrt{3}}\,,
\end{equation}
and its value is restricted by the fact that the
$\mathrm{SU(3)_{L,R}}$ and $\mathrm{U(1)_{X}}$ coupling constants
$g_L=g_R=g$ and $g_X$ must comply with the relation
\begin{equation}
\frac{g_X^2}{g^2}=\frac{\sin^2\theta_W}{1-2(1+\beta^2)\sin^2\theta_W},
\end{equation}
which implies that $\beta^2 <-1+1/(2\sin^2\theta_W)$, hence the
choice $\beta =\sqrt{3}$ is excluded by consistency of the
model~\footnote{Other possibilities such as
  $\beta=\pm\frac{2}{\sqrt{3}}$ and $\beta =0$ will not be considered
  since they lead to exotic fractionary charges for fermions.}.

The spontaneous symmetry breaking of $\mathcal{G}_{3_C 3_L 3_R 1_X}$
is implemented through scalar sextets $\Delta_{L,\hspace{0.3mm}R}$
(will be associated with type-I and type-II seesaw) as well as by the
usual scalar bitriplet $\Phi$ with their matrix representation given
as follows
\begin{eqnarray}
&&\Phi = \begin{pmatrix} \phi^0_{11} & \phi^+_{12} & \phi^{-q}_{13} \\
                         \phi^{-}_{21} & \phi^{0}_{22} & \phi^{-q-1}_{23} \\
                         \phi^{q}_{31} & \phi^{q+1}_{32} & \phi^{0}_{33}
                         \end{pmatrix} \, \sim (\one,\three,\threeS,0)\, , \nonumber \\
&&\rho=\begin{pmatrix}
\rho_{11}^{+}&\rho_{12}^{0}& \rho_{13}^{q+1}\\
\rho_{21}^{0}&\rho_{22}^{-}& \rho_{23}^{q}\\
\rho_{31}^{q+1}& \rho_{32}^{q}& \rho_{33}^{2q+1}
\end{pmatrix} \, \sim(\one ,\three,\three,\frac{2q+1}{3} ) \,, \\
&&\Delta_{L} = \begin{pmatrix} \Delta^0_{11} & \Delta^-_{12}/\sqrt{2} & \Delta^{q}_{13}/\sqrt{2} \\
                         \Delta^{-}_{21}/\sqrt{2} & \Delta^{--}_{22} & \Delta^{q-1}_{23}/\sqrt{2} \\
                         \Delta^{q}_{31}/\sqrt{2} & \Delta^{q-1}_{32}/\sqrt{2} & \Delta^{2q}_{33}
                         \end{pmatrix}_L \, \sim (\one,\six,\one,\frac{2q-2}{3})\, , 
 \nonumber \\
&&\Delta_{R} = \begin{pmatrix} \Delta^0_{11} & \Delta^-_{12}/\sqrt{2} & \Delta^{q}_{13}/\sqrt{2} \\
                         \Delta^{-}_{21}/\sqrt{2} & \Delta^{--}_{22} & \Delta^{q-1}_{23}/\sqrt{2} \\
                         \Delta^{q}_{31}/\sqrt{2} & \Delta^{q-1}_{32}/\sqrt{2} & \Delta^{2q}_{33}
                         \end{pmatrix}_R \, \sim (\one,\one,\six,\frac{2q-2}{3})\, ,  \nonumber \\ 
\end{eqnarray}

We now turn to the Yukawa interactions of the theory. These are
similar to the ones present in the most popular left-right symmetric
models, namely
\begin{equation}
\label{lagyuk}
\begin{split}\mathcal{L}_{y}=&\sum _{\alpha,\beta=1}^{2}\left(h^{Q}_{\alpha\beta}\overline{Q}_{L}^{\,\alpha}\Phi^* Q_{R}^{\beta}\right)+\sum _{\alpha=1}^{2}\left(h^{Q}_{\alpha 3}\overline{Q}_{L}^{\,\alpha}\rho^* Q_{R}^{3}+h^{Q}_{3\alpha}\overline{Q}^{\,3}_{L}\rho Q^\alpha_{R}\right)+h^{Q}_{33}\overline{Q}^{\,3}_{L}\Phi Q^3_{R}+\mathrm{h.c.}
\end{split}
\end{equation}
with $h^Q=(h^Q)^{\dagger}$.  The corresponding spontaneous symmetry
breaking pattern is given by:
\begin{equation}\begin{split}
\vev{\Phi}=\frac{1}{\sqrt{2}}\text{diag}(k_1 ,k_2 ,n)\hspace{2mm} \vev{\Delta_L}&=\frac{1}{\sqrt{2}}\text{diag}(v_L , 0, 0),\hspace{2mm} \vev{\Delta_R}=\frac{1}{\sqrt{2}}\text{diag}(v_R , 0, 0)\,,\\&
\vev{\rho}=\left(\begin{array}{ccc}
0  & k_3 & 0 \\
0  & 0 &0  \\
 0 & 0 &  0\\
\end{array}\right)\,,
\end{split}\end{equation}
Here $\rho$ is the responsible for generating a realistic CKM
matrix. Moreover,
  one assumes, for consistency, that
  $n\,,v_R\gg k_1\,,k_2\,,k_3\,,v_L$.

After spontaneous symmetry breaking, the first line in
Eq.~(\ref{lagyuk}) produces the following Dirac mass matrices for the
Standard Model and exotic quarks:
\begin{equation}\label{mass_up_down}
M^{u}=\frac{1}{\sqrt{2}}\left(
\begin{array}{ccc}
h_{11}^Qk_2 &  h_{12}^Qk_2 & 0 \\
 h_{21}^Qk_2 &  h_{22}^Qk_2 &  0 \\
-h_{31}^Q k_3 &  -h_{32}^Q k_3 &  h_{33}^Qk_1 \\
\end{array}
\right)\,,\qquad 
M^{d}=\frac{1}{\sqrt{2}}\left(
\begin{array}{ccc}
 h_{11}^Qk_1 &  h_{12}^Qk_1 & h_{13}^Q k_3 \\
 h_{21}^Qk_1 &  h_{22}^Qk_1 & h_{23}^Q k_3 \\
0 & 0 &  h_{33}^Qk_2 \\
\end{array}
\right)\,,
\end{equation}
\begin{equation}\label{mass_exotic}
M^{J^{-q-\frac{1}{3}}}=\frac{1}{\sqrt{2}}\left(
\begin{array}{cc}
 h_{11}^Qn &  h_{12}^Qn \\
 h_{21}^Qn &  h_{22}^Qn \\
\end{array}
\right)\,,\qquad  M^{J^{q+\frac{2}{3}}}=h_{33}^Qn.
\end{equation}
\newline
For leptons we have:
\begin{equation}
m_{ab}^\ell=\frac{k_2}{\sqrt{2}}h^\ell_{ab}\hspace{1mm},
\end{equation}
and the new leptons $\chi^q_{L,R}$ form heavy Dirac pairs with masses
\begin{equation}\label{exotic_charged_lepton}
m_{ab}^\chi=\frac{n}{\sqrt{2}}h^\ell_{ab}\hspace{1mm}.
\end{equation}
Concerning neutrinos, the structure of their mass matrix will involve
crucially the Higgs sextets, as explained below.

\subsection{Model with sextets and seesaw mechanism}
\label{sec:model-with-sextets}

We now turn to the Yukawa interactions for neutrinos. These are
similar to the ones present in the most popular left-right symmetric
models, namely
\begin{equation}
\label{lagyuk}
\begin{split}\mathcal{L}_{y}=& \sum_{a,b=1}^{3} \left[ h^{\ell}_{ab}\overline{\psi}_{L}^{\,a}\Phi \psi_{R}^{b}+f_{ab}\left(\overline{\psi}_{L}^{\,c\, a}\Delta_{L}^\dagger\psi_{L}^{b}+\overline{\psi}_{R}^{\,c\, a}\Delta_{R}^\dagger\psi_{R}^{b}\right)\right]+\mathrm{h.c.}
\end{split}
\end{equation}
where $h^{\ell}=(h^{\ell})^\dagger$ and $\Delta_{L,R}$ denote the Higgs
sextets. As a result the neutrino mass matrix can be written as
\begin{equation}
m_{\nu}=\left(\begin{array}{ccc}
M_L  & m_{D} \\
m_{D}^T & M_R \\
\end{array}\right)
\hspace{0.5mm},
\end{equation}
where
\begin{equation}
M_L=2f_{ab}v_L,\hspace{2mm}m_{D}=h^\ell_{ab}k_1, \hspace{2mm}M_R=2f_{ab}v_R\hspace{0.3mm}.
\end{equation}
Thus we obtain the standard combination of type I and
  type II seesaw mechanisms~\cite{Schechter:1980gr}:
\begin{equation}
m_1\approx M_L -m_{D} M_R^{-1} m_{D}^T\,,\qquad m_2 \approx M_R\,.
\end{equation}
for which the diagonalizing matrices can be obtained systematically in
perturbation theory as in~\cite{Schechter:1981cv}.

\ignore{ 
\subsection{\textcolor{red}{Model with triplets}}
\label{sec:model-with-triplets}

\textcolor{red}{In the case of scalar triplets, $\Psi_{L,R}$, Yukawa
  interaction are:
\begin{equation}
-\mathcal{L}_{\text{leptons}} =\sum_{a,b}\left[y_R^{ab} \overline{\psi}^{a}_R \Psi_R S^b_L +y_L^{ab} \overline{\psi}^{a}_L \Psi_L (S^{b}_L)^c  +  y^{ab}\overline{\psi}_{L}^{a}\Phi \psi_{R}^{b} +\frac{\mu}{2}\overline{(S^a_L)^c} S_L^b \right]+\text{h.c.}
\end{equation}
\textcolor{blue}{and the mass matrix
  reads~\cite{Akhmedov:1995ip,Akhmedov:1995vm}}:
\begin{equation}
M_\nu=\left(\begin{array}{ccc}
0  & y k_1  & y_L v_L \\
y^T k_1  & 0 & y_R v_R \\
y_L^T v_L & y_R^T v_R & \mu \\
\end{array}\right).
\end{equation}
\textcolor{blue}{For example, setting $v_L\to 0$ gives
  us~\cite{Mohapatra:1986bd}}
\begin{equation}
m_{light}\approx\frac{k_1^2}{v_R^2} (y\,y_R^{T-1})\mu (y_R^{-1}\,y^T)\,,
\end{equation}
which vanishes as $\mu \to 0$, reproducing the standard inverse seesaw
relation.
\textcolor{blue}{On the other hand, taking $\mu = 0$ leads
  to~\cite{Malinsky:2005bi}}
\begin{equation}
m_{light}\approx\frac{v_L k_1}{v_R} \left[y(y_R\,y_L^{-1})^T+ (y_R\,y_L^{-1}) y^T)\right]\,,
\end{equation}
recovering the linear seesaw mass relation.  Now we will study two
different models based in $\beta=\pm\frac{1}{\sqrt{3}}$ that will
slightly differ in the physics associated to the $SU(3)$ group.}

} 

\subsection{The SVS model: $\beta=\frac{-1}{\sqrt{3}}$}
\label{sec:svs-model:-beta=frac}

In this section we will fix the $\beta$ parameter (see equation
\ref{cheqn}) to the value $\beta =-\frac{1}{\sqrt{3}}$ which
correspond to the SVS model~\cite{Singer:1980sw}.  Notice that in this
case the electric charge of the third component of the leptonic
triplet is $q=0$.  In general the electric charge relation is given
by the following formula
\begin{equation}
	Q = T_{3L} + T_{3R} -\frac{1}{\sqrt{3}} \left( T_{8L} + T_{8R} \right) + X\,.
\end{equation}
\begin{eqnarray}
  \Psi_{aL} &= \begin{pmatrix}\nu_{aL} \\ \ell^-_{aL} \\ N_{aL} \end{pmatrix} \, \sim
  (\one,\three,\one,\frac{-1}{3})\, , & 
                                        \Psi_{aR} = \begin{pmatrix}\nu_{aR} \\ \ell^-_{aR} \\ N_{aR} \end{pmatrix} \, \sim
  (\one,\one,\three,\frac{-1}{3})\, ,\nonumber \\[1mm]
  Q_{\alpha L} &= \begin{pmatrix} d_{\alpha L} \\ u_{\alpha L} \\ D_{\alpha L} \end{pmatrix} \, \sim
  (\three,\threeS,\one,0)\, , & 
                                Q_{\alpha R} = \begin{pmatrix} d_{\alpha R} \\ u_{\alpha R} \\ D_{\alpha R} \end{pmatrix} \, \sim
  (\three,\one,\threeS,0)\,, \nonumber \\[1mm]
  Q_{3 L} &= \begin{pmatrix} u_{3 L} \\ d_{3 L} \\ U_{3 L} \end{pmatrix} \, \sim
  (\three,\three,\mathbf{1},\frac{1}{3})\, , & 
                                               Q_{3 R} = \begin{pmatrix} u_{3 R} \\ d_{3 R} \\ U_{3 R} \end{pmatrix} \, \sim
  (\three,\one,\three,\frac{1}{3})\,.
\end{eqnarray}

The spontaneous symmetry breaking pattern is now more general, since
there are more scalars that can develop a vev. We will assume the
simplest case, e.g. only $\Delta_{33}$ for $\Delta_{L,R}$ will develop
a non-zero vev in order to give Majorana masses to exotic neutrinos:
\begin{equation}
\vev{\Delta_L}=\frac{1}{\sqrt{2}}\text{diag}(v_L , 0, \Lambda_L),\hspace{2mm} \vev{\Delta_R}=\frac{1}{\sqrt{2}}\text{diag}(v_R , 0, \Lambda_R)\,,
\end{equation}
Let us also assume $\Lambda_R\sim v_R$ and $\Lambda_L\sim k_1$. 
In this case the neutrino mass matrix, in the basis
$(\nu_L,\nu_R,N_L,N_R)$, is given by:
\begin{equation}
m_{\nu}=\left(\begin{array}{cccc}
M_L  & m_{D} & 0 & 0 \\
m_{D}^T & M_R& 0 & 0 \\
0 & 0 & M_L^\prime & M_D \\
 0 & 0 & M_D & M_R^\prime \\
\end{array}\right)
\hspace{0.5mm},
\end{equation}
With:
\begin{equation}\begin{split}
&M_L=2f_{ab}v_L,\hspace{2mm}m_{D}=h^\ell_{ab}k_1, \hspace{2mm}M_R=2f_{ab}v_R\hspace{0.3mm},\\&
M_L^\prime=2f_{ab}\Lambda_L,\hspace{2mm}M_{D}=h^\ell_{ab}n, \hspace{2mm}M_R^\prime=2f_{ab}v_R\hspace{0.3mm}.
\end{split}\end{equation}
We also note that in Eq.(\ref{mass_exotic}) there are two exotic
vector-like down-type quarks and an exotic vector-like up-type quark.

\ignore{ 
\subsection{\textcolor{red}{The \"Ozer model: $\beta=\frac{+1}{\sqrt{3}}$}}

\textcolor{red}{The electric charge relation is given by the following formula
\begin{equation}
	Q = T_{3L} + T_{3R} + \frac{1}{\sqrt{3}} \left( T_{8L} + T_{8R} \right) + X\,.
\end{equation}
\begin{eqnarray}
 \Psi_{aL} &= \begin{pmatrix}\nu_{aL} \\ \ell^-_{aL} \\ E^-_{aL} \end{pmatrix} \, \sim
(\one,\three,\one,\frac{-4}{3})\, , & 
\Psi_{aR} = \begin{pmatrix}\nu_{aR} \\ \ell^-_{aR} \\ E^-_{aR} \end{pmatrix} \, \sim
(\one,\one,\three,\frac{-4}{3})\, ,\nonumber \\[1mm]
 Q_{\alpha L} &= \begin{pmatrix} d_{\alpha L} \\ u_{\alpha L} \\ U_{\alpha L} \end{pmatrix} \, \sim
(\three,\threeS,\one,\frac{1}{3})\, , & 
Q_{\alpha R} = \begin{pmatrix} d_{\alpha R} \\ u_{\alpha R} \\ U_{\alpha R} \end{pmatrix} \, \sim
(\three,\one,\threeS,\frac{1}{3})\,, \nonumber \\[1mm]
 Q_{3 L} &= \begin{pmatrix} u_{3 L} \\ d_{3 L} \\ D_{3 L} \end{pmatrix} \, \sim
(\three,\three,\mathbf{1},0)\, , & 
Q_{3 R} = \begin{pmatrix} u_{3 R} \\ d_{3 R} \\ D_{3 R} \end{pmatrix} \, \sim
(\three,\one,\three,0)\,.
\end{eqnarray}
According to leptons this case simpler since exotic 331 leptons are
charged and, with our SSB pattern, will not mix with the SM
ones. Their masses are given by \ref{exotic_charged_lepton}.  The
exotic quark spectrum (see \ref{mass_exotic}) is given by two exotic
up-type quarks and an extra down-type quark.}

} 
\section{Gauge Coupling Unification}
\label{sec:renorm-group-equat-1}

It is interesting to note that the spontaneous symmetry breaking of
the $\mathcal{G}_{3331}$ model to the low energy theory can be
implemented through three possible ways
\begin{align}
      \label{eq:BreakingChain}
      {\bf A:}\quad & \mathbb{G}_{}\, \, \mathop{\longrightarrow}^{M_U} \mathcal{G}_{ 3_C 3_L 3_R 1_{X} }\, 
      \mathop{\longrightarrow}^{n \simeq v_R \simeq \Lambda_R}_{} \mathcal{G}_{3_C 2_L 1_Y}\, 
      \mathop{\longrightarrow}^{k_{i}} \mathcal{G}_{3_C 1_{Q}}\,, \nonumber \\
      {\bf B:}\quad & \mathbb{G}_{}\, \, \mathop{\longrightarrow}^{M_U} \mathcal{G}_{ 3_C 3_L 3_R 1_{X} }\, 
      \mathop{\longrightarrow}^{v_R} \mathcal{G}_{ 3_C 3_L 1_{X^\prime}1_{X^{\prime\prime}} }\,
      \mathop{\longrightarrow}^{n}_{} \mathcal{G}_{3_C 2_L 1_Y}\, 
      \mathop{\longrightarrow}^{k_i} \mathcal{G}_{3_C 1_{Q}}\,, \nonumber \\
      {\bf C:}\quad & \mathbb{G}_{}\, \, \mathop{\longrightarrow}^{M_U} \mathcal{G}_{ 3_C 3_L 3_R 1_{X} }\, 
      \mathop{\longrightarrow}_{}^{n} \mathcal{G}_{ 3_C 2_L 2_R 1_{X} }\,
      \mathop{\longrightarrow}_{}^{v_R\simeq\Lambda_R} \mathcal{G}_{3_C 2_L 1_Y}\, 
      \mathop{\longrightarrow}_{}^{k_i} \mathcal{G}_{3_C 1_{Q}}\,, 
\end{align}
In this work we will focus only on the renormalization group study of
Case-A, and leave the detailed analysis of cases -B and -C for a
follow up study.  The symmetry breaking chain that we will consider is
as follows
\begin{eqnarray*}
&
SU(3)_{C}^{}\times SU(3)_{L}^{}\times SU(3)_{R}^{}\times U(1)_{X}^{} &\nonumber\\
&
\hspace*{-0.8cm} \downarrow \hspace*{-0.0cm} n \simeq v_R \simeq\Lambda_R &\nonumber\\
&
\hspace*{-0.5cm} SU(3)_C^{} \times SU(2)_{L}^{}\times U(1)_{Y}^{} \nonumber\\
&
\hspace*{0.8cm}\downarrow\langle \phi(2,\frac{1}{2},1) \rangle \subset \Phi  &\nonumber\\
&\hspace*{-1.0cm} SU(3)_C^{}  \times U(1)_{Q}^{}\,.&
\end{eqnarray*}
Without presuming any underlying group for grand unification we will
first study the RGEs in this section to explore whether unification of
the three gauge couplings can be obtained in the \3331 theory at a
certain scale $M_U$. Using the RGEs we express the hypercharge (and X)
normalization and the unification scale as a function of \3331
breaking scale. Next we study the allowed range of \3331 breaking
scale such that one can obtain a guaranteed unification of the gauge
couplings. First we discuss the case of the minimal models discussed
in section \ref{sec:svs-model:-beta=frac}. Then, we study the impact
of adding three generations of leptonic octet representations
$(1,8_{L,R},0)$ that can give gauge coupling unification for a TeV
scale \3331 breaking.

The evolution for running coupling constants at one loop level is
governed by the RGEs
\begin{equation}{\label{4.1}}
\mu\,\frac{\partial g_{i}}{\partial \mu}=\frac{b_i}{16 \pi^2} g^{3}_{i},
\end{equation}
which can be written in the form
\begin{equation}{\label{4.2}}
\frac{1}{\alpha_{i}(\mu_{2})}=\frac{1}{\alpha_{i}(\mu_{1})}-\frac{b_{i}}{2\pi} \ln \left( \frac{\mu_2}{\mu_1}\right),
\end{equation}
where $\alpha_{i}=g_{i}^{2}/4\pi$ is the fine structure constant for
$i$--th gauge group, $\mu_1, \mu_2$ are the energy scales with
$\mu_2 > \mu_1$. The beta-coefficients $b_i$ determining the evolution
of gauge couplings at one-loop order are given by
\begin{eqnarray}{\label{4.3}}
	&&b_i= - \frac{11}{3} \mathcal{C}_{2}(G) 
				 + \frac{2}{3} \,\sum_{R_f} T(R_f) \prod_{j \neq i} d_j(R_f) 
  + \frac{1}{3} \sum_{R_s} T(R_s) \prod_{j \neq i} d_j(R_s).
\label{oneloop_bi}
\end{eqnarray}
Here, $\mathcal{C}_2(G)$ is the quadratic Casimir operator for the
gauge bosons in their adjoint representation,
\begin{equation}{\label{4.4}}
	\mathcal{C}_2(G) \equiv \left\{
	\begin{matrix}
		N & \text{if } SU(N), \\
    0 & \text{if }  U(1).
	\end{matrix}\right.
\end{equation}
On the other hand, $T(R_f)$ and $T(R_s)$ are the Dynkin indices of the
irreducible representation $R_{f,s}$ for a given fermion and scalar,
respectively, 
\begin{equation}{\label{4.5}}
	T(R_{f,s}) \equiv \left\{
	\begin{matrix}
		1/2 & \text{if } R_{f,s} \text{ is fundamental}, \\
    N   & \text{if } R_{f,s} \text{ is adjoint}, \\
		0   & \text{if } R_{f,s} \text{ is singlet},
	\end{matrix}\right.
\end{equation}
and $d(R_{f,s})$ is the dimension of a given representation $R_{f,s}$
under all gauge groups except the $i$-th~gauge group under
consideration. An additional factor of $1/2$ is multiplied in the case
of a real Higgs representation.

The charge equation is given in Eq. (\ref{cheqn}), where the
generators (Gell-Mann matrices) are normalized as
$\text{Tr} (T_{i}T_{j})=\frac{1}{2}\delta_{ij}$. We define the
normalized hypercharge operator $Y_{N}$ and $X_{N}$ as\\
\begin{equation}{\label{4.7}}
Y=n_{Y} Y_{N},\quad X=n_{X}X_{N},
\end{equation}
such that we have\\
\begin{equation}{\label{4.8}}
n_{Y}^{2}=(1+2\beta^{2})+n_{X}^{2},
\end{equation}
and the normalized couplings are related by\\
\begin{equation}{\label{4.9}}
n_{Y}^{2} {\left(\alpha^{N}_{Y}\right)}^{-1}=\beta^{2}\alpha_{3L}^{-1}+(1+\beta^{2})\alpha_{3R}^{-1}+\left[n_{Y}^{2}-(1+2\beta^{2})\right]{\left(\alpha^{N}_{X}\right)}^{-1},
\end{equation}
at the \3331 symmetry breaking scale, where\\
\begin{equation}{\label{4.10}}
\alpha^{N}_{Y} =n_{Y}^{2} \alpha_{Y}, \quad \alpha^{N}_{X}=\left[n_{Y}^{2}-(1+2\beta^{2})\right]\alpha_{X}, \quad \alpha_{3L}=\alpha_{2L}.
\end{equation}
Furthermore, to keep things simple and minimal we will assume
$\alpha_{3L}=\alpha_{3R}$ and the left-right symmetry of the \3331
model ensures that $b_{3L}=b_{3R}$. The RG running for the phase
between the electroweak symmetry breaking and the \3331 symmetry
breaking is described by the the SU(3)$_C$ coefficient $b_{s}$, the
SU(2)$_L$ coefficient $b_{2L}$ and the U(1)$_Y$ unnormalized
coefficient $b^{\text{UN}}_{Y}$. Likewise, in the unbroken \3331
phase, the running coefficients for the SU(3)$_C$, SU(3)$_L$,
SU(3)$_R$ and unnormalized U(1)$_X$ components are $b^{X}_{3C}$,
$b_{3L}$, $b_{3R}$ and $b^{\text{UN}}_{X}$, respectively. The scale
$M_{Z}$ corresponds to the $Z$ boson-pole, the \3331 symmetry breaking
scale is denoted by $M_{X}$ and $M_{U}$ is the scale of unification
for the normalized gauge couplings. From the above set of equations
the unification scale $M_U$ and $n_{Y}^{2}$ can be expressed as a
function of $M_{X}$.

\subsection{The minimal SVS Model with sextet Higgs sector}
\label{sec:minimal-svs-model}

The first case of interest is the minimal scenario described in
section \ref{sec:svs-model:-beta=frac}. The relevant gauge quantum numbers for the fermions and the scalars relevant for the RG running of the beta-coefficients in different phases have been tabulated in Table \ref{tab:bi_DPVatMP_MwMz_triplets}. For the phase between the
electroweak symmetry breaking and the \3331 symmetry breaking the
one-loop beta-coefficients are given by $b_{2L} = -19/6$,
$b^{\text{UN}}_{Y} = 41/6$, $b_{3C} = -7$, while for the phase between
the \3331 symmetry breaking and gauge coupling unification the
one-loop beta-coefficients are given by $b_{3L} = -31/6$,
$b^{\text{UN}}_{X} = 43/9$, $b^{X}_{3C} = -5$.

\begin{table}[h!]
\begin{center}
\begin{tabular}{|c|c|c|}
\hline
Group $G_{I}$  & Fermions      & Scalars            \\
\hline \hline
$\begin{array}{l}
\mathcal{G}_{321} \\
(M_Z \leftrightarrow v_R)  \end{array}$  & 
$\begin{array}{l}
Q_{aL}(3,2,1/6) \\
u_{aR}(3,1,2/3), d_{aR}(3,1,-1/3) \\
\ell_{aL}(1,2,-1/2), e_{aR}(1,2,-1)
  \end{array}$ &
$\begin{array}{l}
\phi(1,2, \frac{1}{2}) \end{array}$
 \\
\hline 
$\begin{array}{l}
{\small G_{3331}}\\
(v_R \leftrightarrow M_U)  \end{array}$  & 
${\small \begin{array}{l}
\Psi_{aL}(1,3,1,\frac{q-1}{3}), \Psi_{aR}(1,1,3,\frac{q-1}{3}) \\
Q_{\alpha L}(3,3,1,\frac{-q}{3}), Q_{\alpha R}(3,1,3,\frac{-q}{3})  \\
Q_{3 L}(3,3,1,\frac{q+1}{3}), Q_{3 R}(3,1,3,\frac{q+1}{3})  \end{array}}$ &
${\small \begin{array}{l}
\Phi(1,3,3^*,0) \\ 
\rho(1,3,3,\frac{2q+1}{3}) \\
\Delta_L(1,6,1,\frac{2(q-1)}{3}) \\
\Delta_R(1,1,6,\frac{2(q-1)}{3}) \end{array}}$                                                                                       
\\
\hline
\hline
\end{tabular}
\caption{Table showing the gauge quantum numbers for the fermions and
  the scalars, and the beta-coefficients for the renormalization group
  evolution in different phases of gauge symmetry.}
\label{tab:bi_DPVatMP_MwMz_triplets}
\end{center}
\end{table}

\begin{figure}[t!]
\includegraphics[width=0.47\linewidth]{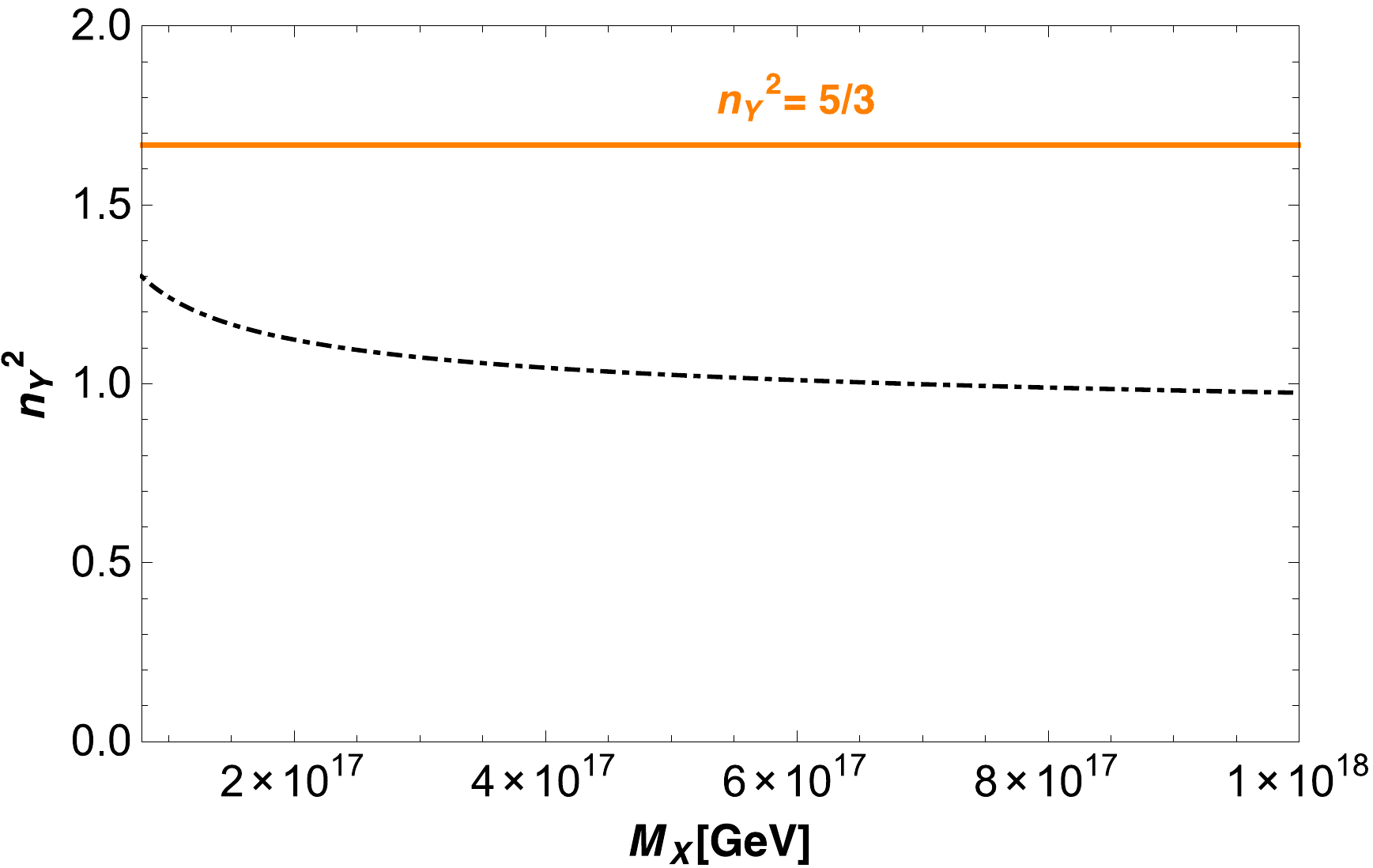}
\caption{The hypercharge normalization factor $n_{Y}^{2}$ as a
  function of $M_{X}$, the scale of \3331 symmetry breaking, such that
  there is a guaranteed unification at some scale $M_U$:
  $M_{X}\leq M_{U}\leq 10^{18}$~GeV. The solid line gives $n_{Y}^{2}$
  as a function of $M_X$ and the dot-dashed line shows the lower
  limit $n_{Y}^{2}= \frac{5}{3}$ for the allowed value of $n_{Y}^{2}$
  from Eq. (\ref{4.8}).}
\label{fig1}
\end{figure}

As shown in Fig.~\ref{fig1} we plot the hypercharge normalization factor
$n_{Y}^{2}$ as a function of \3331 symmetry breaking scale $M_{X}$
such that there is a guaranteed unification. We have chosen the $M_X$
scale such that there is a guaranteed unification at some scale $M_U$:
$M_{X}\leq M_{U}\leq 10^{18}$~GeV. The dashed horizontal red line
represents the lower limit $n_{Y}^{2}= \frac{5}{3}$ for the allowed
value of $n_{Y}^{2}$ from eq.~(\ref{4.8}). Interestingly,
$n_{Y}^{2}= \frac{5}{3}$ is also the standard $SU(5)$
normalization. It can be seen from the figure, for the $M_X$ range
allowed by the condition that there is a guaranteed unification at
some scale $M_U$: $M_{X}\leq M_{U}\leq 10^{18}$~GeV, the hypercharge
normalization $n_{Y}^{2}$ is almost constant $\approx 1.3$ and is
clearly below the allowed lower limit (implying a negative
$n_{X}^{2}$, which is unphysical). Thus it is not possible to obtain
gauge coupling unification for the minimal scenario described in
section-\ref{sec:model-framework}. Interestingly, we find that adding
sequential left- and right-handed fermionic octets for each generation
it is possible to obtain a consistent unification for this model while
having the \3331 scale and the octet scale within the reach of the
LHC.

\subsection{The SVS Model with fermionic octets}
\label{sec:svs-model-with}

In addition to the field content discussed above, we now include three
generations of fermion octets $\Omega_{L,R}$ with the assignments
under the \3331 group given by
\begin{equation}{\label{4.15}}
 \Omega_{L}\equiv [1, 8, 1, 0], \quad \Omega_{R}\equiv [1, 1, 8, 0]. 
\end{equation}

\begin{figure}[t!]
\includegraphics[width=0.49\linewidth]{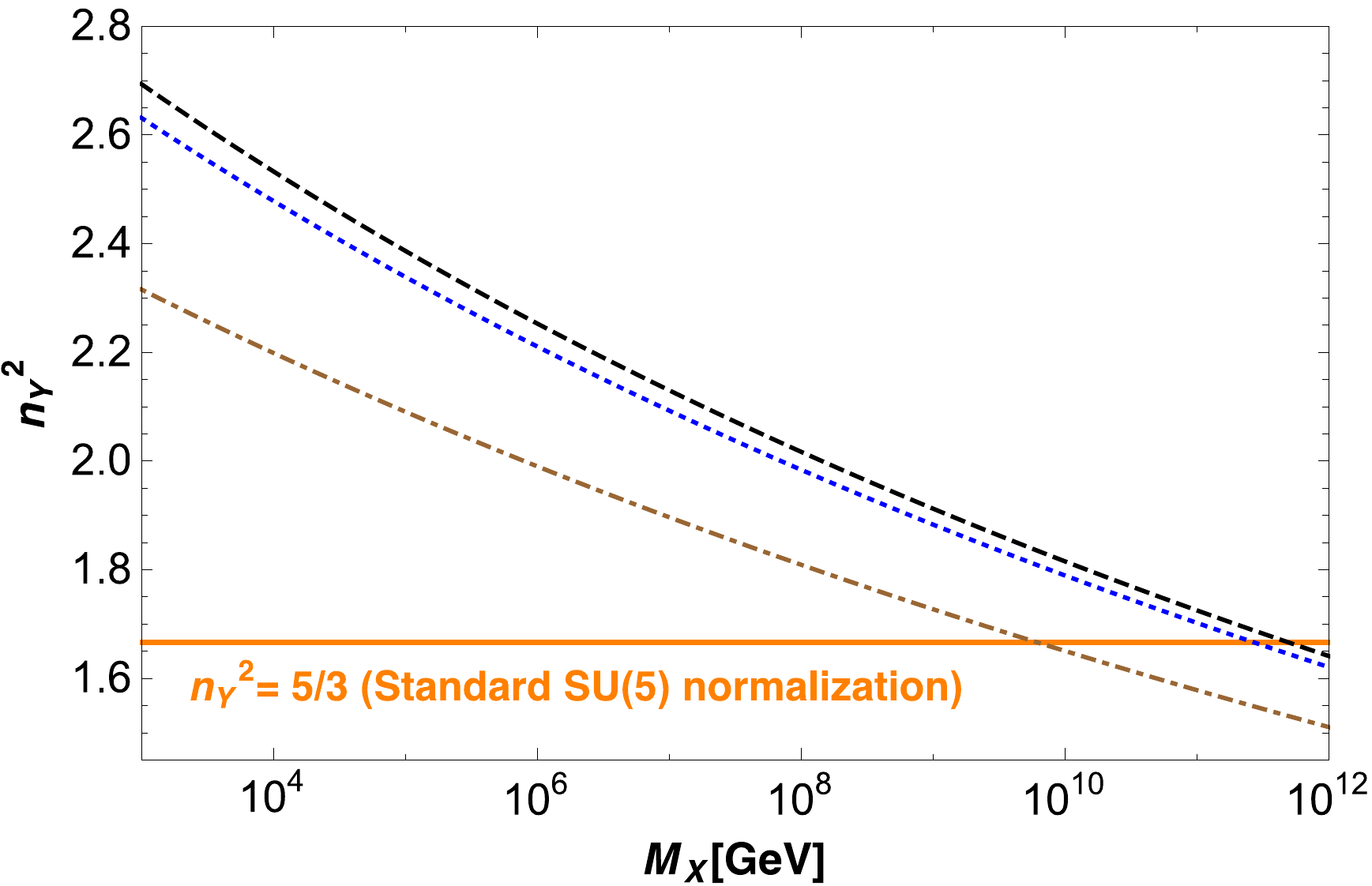}
\includegraphics[width=0.49\linewidth]{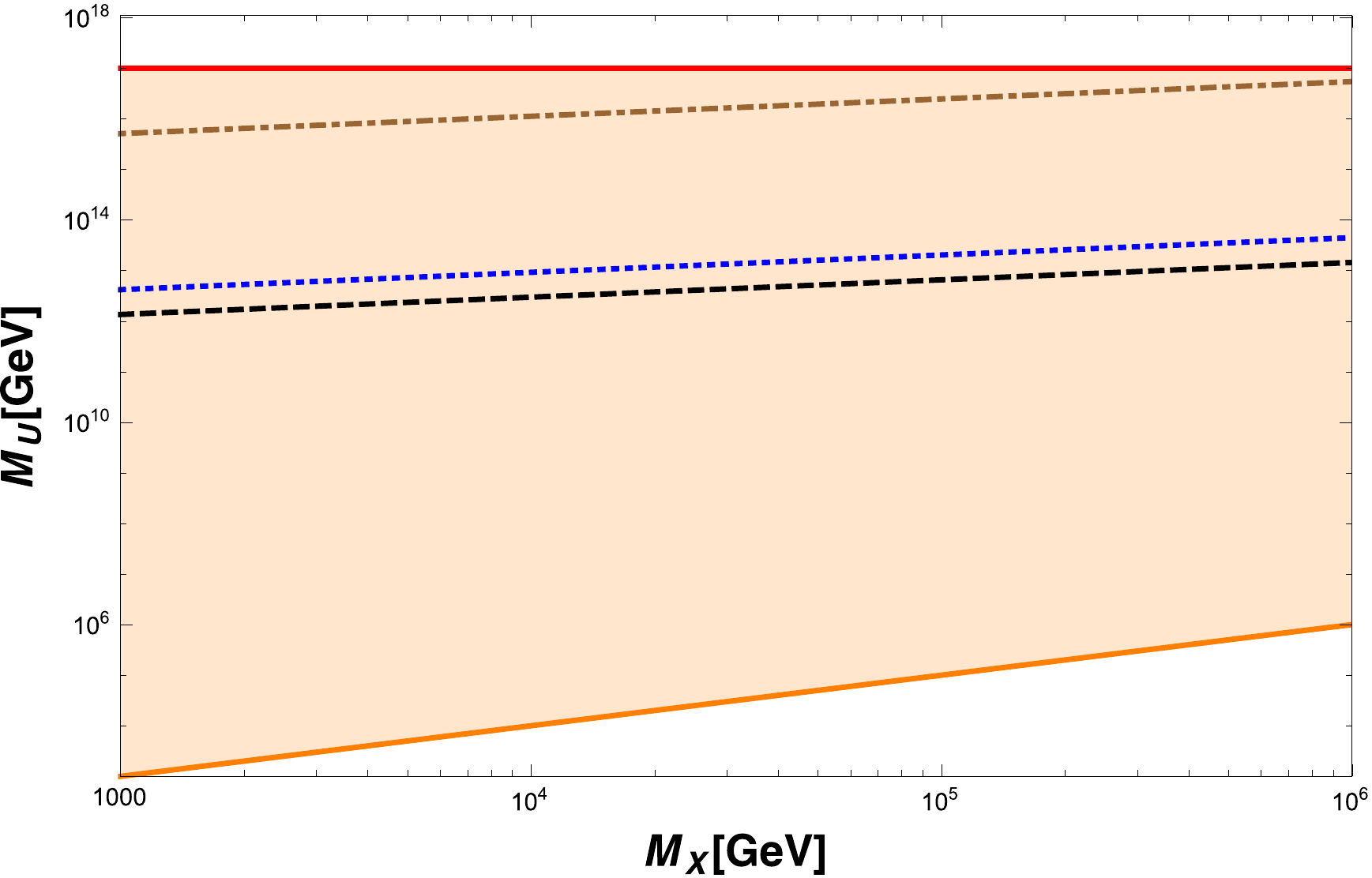}
\caption{(left) Plot showing the hypercharge
  normalization factor $n_{Y}^{2}$ as a function of the \3331 symmetry
  breaking scale $M_{X}$ such that there is a guaranteed unification
  at some scale $M_U$: $M_{X}\leq M_{U}\leq 10^{18}$~GeV. The black
  dashed, blue dotted, and brown dot-dashed curves correspond to the
  cases $M_8/M_X= 1, 3$, and $3\times 10^{3}$ respectively. The orange
  solid line shows the lower limit $n_{Y}^{2}= \frac{5}{3}$ for the
  allowed value of $n_{Y}^{2}$ from Eq. (\ref{4.8}) which is also the
  standard SU(5) normalization.
  (right) Plot showing the allowed range for $M_X$ for which
  unification is guaranteed at a scale
  $M_{X}\leq M_{U}\leq 10^{18}$~GeV. The orange lower boundary
  corresponds to $M_{X}\leq M_{U}$ and the upper red boundary
  corresponds to $10^{18}$~GeV. The black dashed, blue dotted, and
  brown dot-dashed curves correspond to the cases $M_8/M_X= 1, 3$, and
  $3\times 10^{3}$ respectively. }
\label{fig7}
\end{figure}

In order to make the renormalization group evolution analysis general
we will keep the octet mass scale $M_8$ as a separate scale from the
\3331 symmetry breaking scale and assume it to lie somewhere in
between the \3331 symmetry breaking scale and the unification
scale. For the phases between the \sm and \3331 symmetry breaking
scale, and the \3331 symmetry breaking and the octet scale $M_8$ the
one-loop beta-coefficients remain the same as discussed in the
previous section. For the phase between the octet scale $M_8$ and
gauge coupling unification the one-loop beta-coefficients are given by
$b_{3L} = 5/6$, $b^{\text{UN}}_{X} = 43/9$, $b^{X}_{3C} = -5$. In
Fig. \ref{fig7} (right) we plot the allowed range for $M_X$ for which
unification is guaranteed at a scale $M_{X}\leq M_{U}\leq 10^{18}$GeV.
Interestingly, in this scenario we find that for a \3331 symmetry
breaking scale $M_{X}$ as low as a few TeV it is possible to achieve
unification \footnote{ We should emphasize that we here do not specify
  the GUT group and thus the limit coming from the non-observation of
  proton decay is beyond the scope of the current discussion.}. We
will show the plot for three distinct and physically interesting
values of the ratio of the octet mass scale to the \3331 symmetry
breaking scale $M_8/M_X$. In Fig. \ref{fig7} (left) we plot the
hypercharge normalization factor $n_{Y}^{2}$ as a function of the
\3331 symmetry breaking scale $M_{X}$ such that there is a guaranteed
unification at some scale $M_U$: $M_{X}\leq M_{U}\leq
10^{18}$~GeV.
The black dashed, blue dotted, and brown dot-dashed curves correspond
to the cases $M_8/M_X= 1, 3$, and $3\times 10^{3}$ respectively. The
orange solid line shows the lower limit $n_{Y}^{2}= \frac{5}{3}$ for
the allowed value of $n_{Y}^{2}$ from Eq. (\ref{4.8}) which is also
the standard $SU(5)$ normalization. Note that in this case one can
have a TeV scale \3331 symmetry breaking and octet mass scale
consistent with unification with the hypercharge normalization
$n_{Y}^{2}$ is well above the allowed lower limit. This justifies the
addition of the octets to the spectrum and makes the scenario testable
in the current and near future collider experiments.

\begin{figure}[t!]
\includegraphics[width=0.7\linewidth]{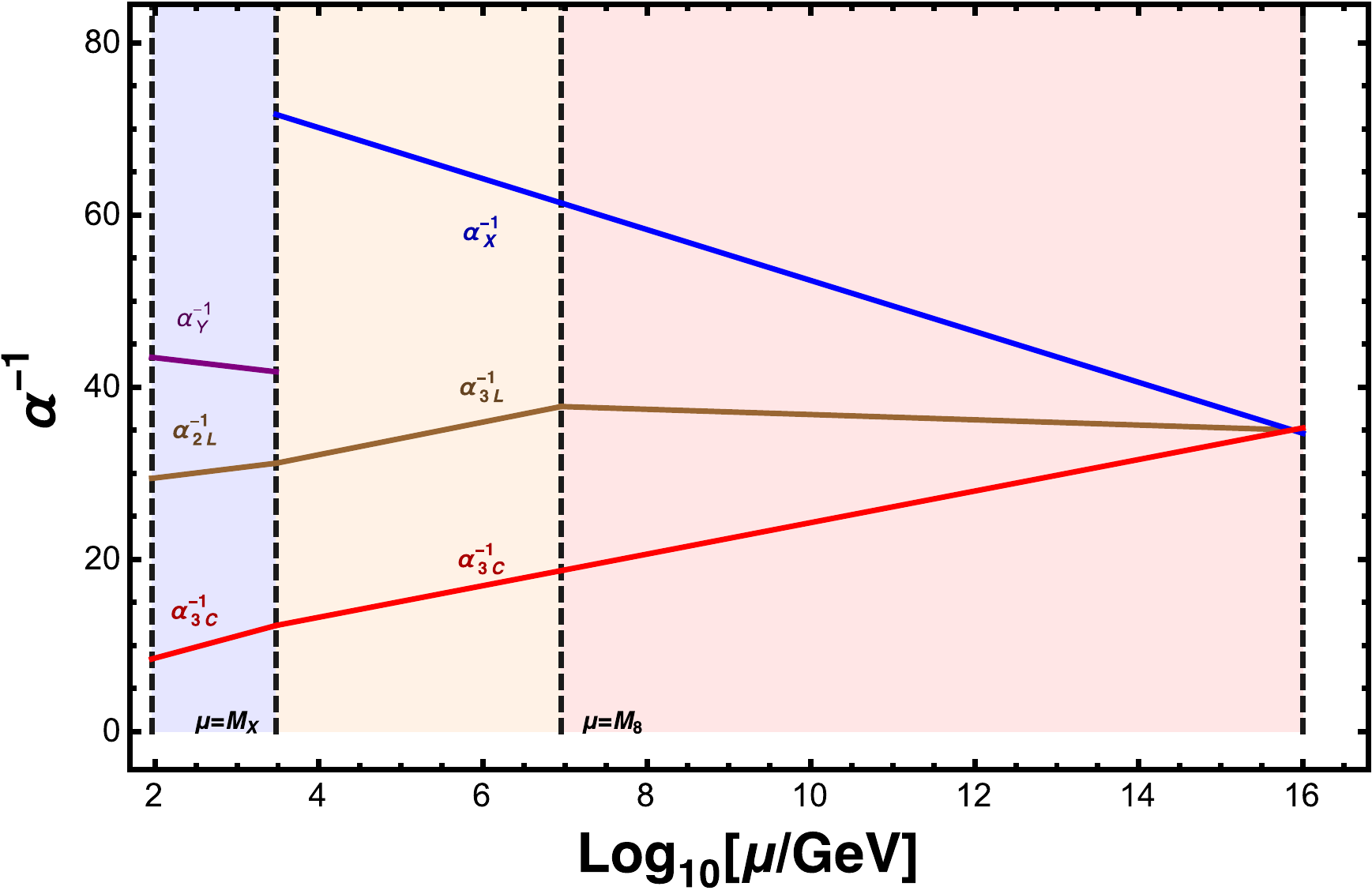}
\caption{Gauge coupling running in the SVS Model adding three
  generations of leptonic octets with \3331 symmetry breaking scale at
  $M_{X}= 3000$~GeV, and the octet mass scale at
  $M_8=3\times 10^{3} M_X$ demonstrating successful gauge unification
  at the scale $M_{U} = 10^{16}$~GeV with $n_{Y}^{2}=2.26$.}
\label{fig9}
\end{figure}

In Fig.~\ref{fig9} we show an example of gauge coupling evolution
with the \3331 symmetry breaking scale at $M_{X}= 3000$~GeV, and the
octet mass scale at $M_8=3\times 10^{3} M_X$ demonstrating successful
gauge coupling unification at a scale $M_{U} = 10^{16}$~GeV with
$n_{Y}^{2}=2.26$. Thus, from the perspective of a low \3331 symmetry
breaking scale within the reach of accelerator experiments like the
LHC $\sim{\mathcal{O}}$(TeV)) this model is a very interesting
candidate leading to a successful gauge coupling unification. In
addition to the new gauge bosons, the model can also have a number of
new states associated to the new exotic fermions as well as extra
Higgs bosons which can be searched for at the LHC and other near
future accelerator experiments.

%
\section{Comments on phenomenology}
\label{sec:comm-phen}
In order to achieve gauge coupling unification at a reasonable scale
with the \3331 breaking scale around the few TeV scale the leptonic
octets play a crucial role. Since they are in the $SU(3)$ adjoint
representation they introduce no anomalies, but because of group
theoretic considerations they can produce, though in a minor way, changes in the
relevant mass matrices. From SU(3) group theory we know that
\begin{equation}\begin{split}\label{oct_mass}
&\three\times\threeS=\one + \eight\,,\\&
\threeS\times\threeS=\three_a + \sixS_s\,,\\&
\six\times\three=\eight + \mathbf{10}\,,\\&
\six\times\threeS=\three + \mathbf{15}\,,
\end{split}\end{equation}
thus there are some allowed couplings involving the octets $\Omega$:
\begin{equation}\begin{split}
&\three\times\threeS\times\eight\,,\\&
\three\times\six\times\eight\,,\\&
\eight\times\eight\,.
\end{split}\end{equation}
The first term is relevant for the low-scale seesaw version of the
model~\cite{Reig:2016vtf}, while the second is relevant for the
high-scale seesaw case considered here. The third term will be always
present and gives the octets a bare mass denoted by $M_8$ in our
construction (see \ref{sec:svs-model-with}). A more detailed analysis
can show that with our quantum numbers the sextet mass term in
equation \ref{oct_mass} is actually not allowed by hypercharge
conservation, so that the octets decouple from the neutrino mass
matrix. This is in sharp contrast with the scenario considered in
Ref. \cite{Boucenna:2014dia} in the context of the simpler 3-3-1
model, where the octets were responsible for radiative neutrino mass
generation. Thus in our present case the neutrino masses are simply
given by the seesaw mechanism, and the phenomenology remains
unaffected by adding these new leptonic octets. 

Before closing this discussion we note that, as seen above, in the
presence of octets, unification can take place for relatively low
values of the extended electroweak scale. This implies that, say, a
new $Z^\prime$ lying around the few TeV scale should be produced in a
Drell-Yan process and act as a portal to access the messenger of
neutrino mass generation. In the present version of the seesaw
mechanism, this requires small Yukawa couplings in order to get small
neutrino masses. 

Thus in such scenario there would be no measurable \lfv at low
energies, yet sizeable \lfv at the high energies accessible at the LHC
would be expected~\cite{Das:2012ii,Deppisch:2013cya}.
Moreover, such $Z^\prime$ would have flavor changing neutral current
(FCNC) couplings to quarks, leading to $Z^\prime$-induced K, D and B
neutral meson mass differences, and hence to transitions in the
${K^0}-\bar{K^0}$, ${D^0}-\bar{D^0}$ and ${B^0}-\bar{B^0}$ meson
systems.
For a study of complementarity between FCNC and dilepton resonance
searches at the LHC see Ref.~\cite{Queiroz:2016gif}.


\section{Comments on Grand Unification}
\label{sec:comm-grand-unif}
  
In this section we comment on various possible ways of embedding our
model within a grand unification scheme. Notice that the group \3331
has rank 7, the same as the rank of SU(8). However, none of the
subgroups of SU(8) contains three SU(3), hence the minimal SU(N) group
which can embed the \3331 model is actually SU(9). A possible symmetry
breaking chain to to obtain the \3331 model from SU(9) is given by
\begin{equation}
\text{SU(9)}  \to  \text{SU(6)}\times \text{SU(3)} \times U(1)^{\prime}  \to  \text{SU(3)}\times \text{SU(3)}\times \text{SU(3)}\times U(1)\,.
\end{equation}
In general SU(N) groups have anomalies that one must cancel in order
to keep gauge invariance. One can compute \cite{Banks:1976yg} the
contribution of the $m\text{-rank}$ antisymmetric representation of
SU(N) as:
\begin{equation}
\mathcal{A}=(N-2m)\frac{(N-3)!}{(m-1)!(N-m-1)!}\,,
\end{equation}
so that the anomaly contributions for different $\text{SU(9)}$
antisymmetric representations are given by
\begin{equation}
\mathcal{A}[\mathbf{9}]=1\,,\,\,\mathcal{A}[\mathbf{36}]=5\,,\,\,\mathcal{A}[\mathbf{84}]=9\,,\,\,\mathcal{A}[\mathbf{126}]=5\\.
\end{equation}
Note that one can also introduce symmetric representations like the
$\mathbf{45}$, but due to its large anomaly contribution
$\mathcal{A}[\mathbf{45}]=13$ we will not consider it. The anomaly
contribution for larger symmetric representations grows rapidly. When
constructing a Grand Unified Theory one must not only cancel
anomalies, but also to make sure that the theory is able to generate
the observed \sm chirality \cite{Fonseca:2015aoa}. Some anomaly free
sets capable of reproducing \sm chirality are given by
\begin{equation}\begin{split}
&15(\overline{\mathbf{9}})+3(\mathbf{36})\,,\\&
9(\overline{\mathbf{9}})+\mathbf{84}\,,\\&
3(\overline{\mathbf{36}})+3(\mathbf{126})\,,\\&
14(\overline{\mathbf{9}})+2(\mathbf{36})+\mathbf{84}+\overline{\mathbf{126}}\,,\\&
3(\overline{\mathbf{9}})+3(\overline{\mathbf{36}})+2(\mathbf{84})\,,\\&
5(\overline{\mathbf{9}})+\overline{\mathbf{36}}+2(\mathbf{126})\,,\\&
4(\overline{\mathbf{9}})+2(\overline{\mathbf{36}})+\mathbf{84}+\mathbf{126}\,,
\end{split}\end{equation}
We note that these sets contain the \sm fermion content plus
vector-like representations. It is important to notice that
combinations like $5(\mathbf{9})+\overline{\mathbf{36}}$ are anomaly
free but only reproduce one family. This can be seen when decomposing
all given combinations in terms of the $SU(9)$ subgroup: $SU(5)$
\cite{Georgi:1979md}.  
It is also interesting that the combination given by  
\begin{equation}
\mathbf{1}+\overline{\mathbf{9}}+\mathbf{36}+\overline{\mathbf{84}}+\mathbf{126}\,,
\end{equation}
is anomaly free but does not reproduce the required \sm chirality,
since it corresponds to the $\text{SO(18)}$ spinor representation
$\mathbf{256}$. Despite being complex and chiral
\cite{Wilczek:1981iz}, the latter decomposes into a vector-like set of
representations of its subgroups.  The next question that one should
address is how to match acceptable multiplicities of the different
representations of the \3331 model. 
An interesting possibility is to rely upon F-theory GUTs in order to
obtain the required representations with the correct multiplicity
\cite{Beasley:2008kw,King:2010mq,Callaghan:2011jj,Callaghan:2013kaa}.

  An interesting alternative is to use the $\text{SO(18)}$ gauge group
  \cite{GellMann:1980vs,Wilczek:1981iz,Fujimoto:1981bv,Dong:1982wq,BenTov:2015gra}
  which can also embed our \3331 model. A possible symmetry breaking
  chain to the \3331 model from SO(18) is given by
\begin{equation}
\text{SO(18)} \to  \text{SU(9)} \times U(1)^{\prime\prime}   \to  \text{SU(6)}\times \text{SU(3)} \times U(1)^{\prime} \rightarrow \text{SU(3)}\times \text{SU(3)}\times \text{SU(3)}\times U(1)\,.
\end{equation}
The choice of SO(18) is particularly interesting, as it may
potentially unify all of the fermionic fields within a single spinor
representation $\mathbf{256}$. However, the problem of having unwanted
mirror families contained in the $\mathbf{256}$ is still an open
question.
Another interesting choice of gauge group for embedding the \3331
model is an extended Pati-Salam like gauge group
$\text{SU(4)} \times \text{SU(3)} \times \text{SU(3)}$. A detailed
study of the GUT embedding of the \3331 model is beyond the scope of
this article. Here we have taken a phenomenological approach to the
problem, keeping the generator normalization as a free parameter. Once
a particular GUT embedding is chosen and the field content of the
\3331 model is fully specified in correct representations, the
hypercharge normalization can be determined and Fig. \ref{fig7} can be
readily used in order to check whether a few TeV scale \3331 breaking
scale is consistent with the corresponding GUT embedding.

\section{Summary}
\label{sec:summary}

In this work we have considered the possibility of gauge coupling
unification in a simple model realization of the left-right symmetric
$\mathrm{SU(3)_c \times SU(3)_L \times SU(3)_R \times U(1)_{X}}$ gauge
theory.  Our ``bottom-up'' analysis of the renormalization group
equations for the SVS model with sextets (minimal model) shows that
gauge coupling unification with the 3331 scale at $\mathcal{O}$(TeV)
is possible in the presence of leptonic octets. Interestingly, unlike
in the chiral 3-3-1 model, in the minimal \3331 model these octets do
not affect neutrino mass generation. Consequently, the neutrino masses
arise exclusively from the seesaw mechanism in this model.  It is also
interesting that, due to hypercharge normalization requirements, one
can not achieve unification without these octets.  We have also
briefly commented on possible phenomenological implications of this
model and on possible grand unified theory frameworks which can embed
the $\mathrm{SU(3)_c \times SU(3)_L \times SU(3)_R \times U(1)_{X}}$
model as an intermediate symmetry.

\section*{Acknowledgements}

Work supported by Spanish grants FPA2014-58183-P, Multidark
CSD2009-00064, SEV-2014-0398 (MINECO), PROMETEOII/2014/084
(Generalitat Valenciana).  C.A.V-A. acknowledges support from CONACyT
(Mexico), grant 274397.  

\bibliographystyle{apsrev} \providecommand{\url}[1]{\texttt{#1}}
\providecommand{\urlprefix}{URL }
\providecommand{\eprint}[2][]{\url{#2}}
\bibliography{merged_Valle,newrefs,d27,3331,chrefs,z4}
\end{document}